\documentclass[twocolumn,pra,showpacs,amsmath,amssymb]{revtex4-1}
\usepackage[dvipdfmx]{graphicx}
\usepackage{setspace}
\usepackage{color}
\usepackage{ulem} 
\usepackage{here}
\graphicspath{{./figures/}}

\usepackage{amsmath,amsthm,amssymb}
\usepackage{mathrsfs}
\usepackage{braket,comment}
\usepackage{bm}

\newcommand{\hc}{\hat{c}}

\usepackage[colorlinks=true, citecolor=blue, urlcolor=blue, linkcolor=blue,
setpagesize=false, bookmarks=false
]{hyperref}

\begin{document}
\title{Optical response of the tightbinding model on the Fibonacci chain}
\author{Hiroki Iijima}
\author{Yuta Murakami}
\author{Akihisa Koga}
\affiliation{Department of Physics, Tokyo Institute of Technology, Meguro, Tokyo 152-8551, Japan}
\date{\today}

\begin{abstract}
  We theoretically study the optical conductivity of the tightbinding model 
  which has two types of the hopping integrals arranged in the Fibonacci sequence.
  Due to the lack of the translational symmetry,
  many peak structures appear in the optical conductivity as well as the density of states.
  When the ratio of two hopping integrals is large,
  the self-similar structure appears in the optical conductivity.
  This implies that the optical response between the high-energy bands is related to 
  that within the low-energy bands, which should originate from critical behavior in the wave functions.
  The effects of disorders on the optical conductivity are also analyzed
  in order to show the absence of the self-similarity in the tightbinding model with the random sequence. 
\end{abstract}

\maketitle

\section{INTRODUCTION}
Quasicrystal has been attracting much interests since the first discovery of
the quasicrystalline phase in the Al-Mn alloy~\cite{SchehitmanD}.
Experimental and theoretical efforts have been made to explore and understand physical properties
inherent in quasicrystals~\cite{PhysRevLett.55.2883, Tsai_1987, Tsai_1988, 199098, RevModPhys.65.213, Tsai:2000vk, Kamiya:2018vj, PhysRevMaterials.3.061601, KimuraHeatResist, InabaHeatcapacities, Tamura:2021tk}. 
Among them, the Au-Al-Yb alloy with Tsai-type clusters~\cite{Au-Al-Yb}
exhibits interesting properties at low temperatures.
In the quasicrystal Au$_{51}$Al$_{34}$Yb$_{15}$, quantum critical behavior appears,
while heavy fermion behavior appears in the approximant
Au$_{51}$Al$_{35}$Yb$_{14}$~\cite{Deguchi:2012vd}.
This distinct behavior stimulates theoretical investigations
on the quasiperiodic structures in correlated electron
systems~\cite{Takemori, Takemura, Watanabe, Otsuki, Andrade, KogaTsunetsugu, Takemori2020,Sakai2021,Ghadimi2021}.
The optical response characteristic of the quasicrystals have also been observed.
In the Al-Cu-Fe quasicrystal,
the linear-$\omega$ dependence in the optical conductivity, which is distinct from the conventional
impurity scattering, has been observed~\cite{oc_of_QC_Al_Cu_Fe}.
Furthermore, the direction dependent optical conductivity has been reported
in the Al-Co-Cu quasicrytal~\cite{PhysRevLett.72.1937}.
These studies suggest that the exotic optical properties arise from the quasiperiodic structure. 
Although the low-frequency behavior has theoretically been examined for quasiperiodic systems~\cite{PhysRevLett.60.1672, PhysRevB.70.144207,Varma,TsuneII},
the optical response at finite frequencies has not been discussed in detail.
An important point is that the spatial features of the initial and final states play
a crucial role for the optical process.
It is known that spatially extended states characterized by the momenta are realized in the periodic system,
while critical states are, in general, realized in the quasiperiodic one ~\cite{doi:10.1143/JPSJ.55.1420, PhysRevB.34.563,PhysRevB.40.7413, Kohmoto-critical1}.
Therefore, it is instructive to clarify the optical response inherent in quasiperiodic systems 
by comparing them to systems with distinct properties of eigenstates.

Motivated by this, 
we treat the tightbinding model
which has two types of the hopping integrals arranged in the Fibonacci sequence,
as a simple model.
It is known that each eigenstate for the system shows critical behavior~\cite{Kohmoto-critical1, PhysRevB.34.563, PhysRevB.40.7413, KohmotoKadanoff, MaceJagannathan}, which are characterized by multifractal properties and
the power law decay of the amplitude of the wave functions in the real space.
This eigenstate property is distinct from those for the periodic systems
where the spatially extended states are realized.
We then discuss the optical response in the tightbinding model on the Fibonacci chain,
examining the matrix elements of the current operator, 
which play a crucial role for the optical conductivity.

The paper is organized as follows.
In Sec.~\ref{sec:model}, we introduce the tightbinding model on the Fibonacci chain
and derive the expression of the optical conductivity in this system.
In Sec.~\ref{sec:result},
we discuss the optical response inherent in the Fibonacci chain,
comparing with that in the approximants.
The effect of the disorders is also addressed.
A summary is given in the last section.

\section{MODEL AND METHOD}\label{sec:model}
We consider the tightbinding model to study
the optical conductivity inherent in the Fibonacci chain.
The Hamiltonian is given as
\begin{align}
  \label{eq:hamiltonian}
  \hat{H}(t) &=- \sum_n\left( v_n e^{-iq L_n A(t)} \hc_n^{\dag} \hc_{n+1} + \mathrm{H.c.}\right),
\end{align}
where $\hc_n (\hc_n^{\dag})$ is the annihilation (creation) operator of
a spinless fermion at the $n$th site,
$v_n$ ($L_n$) denotes the hopping integral (lattice spacing)
between the $n$th and $(n+1)$th sites,
 and $q$ is the charge of the fermion.
The site-independent vector potential $A(t)$ leads to
the uniform electric field ${\cal E}(t)= - \partial_t A(t)$.

Here, we introduce the Fibonacci sequence ${\cal S}$
into the Hamiltonian.
It is known that the Fibonacci sequence is generated
by means of the substitution rule for two letters $L$ and $S$:
$L\rightarrow LS$ and $S\rightarrow L$.
Applying the substitution rule to the initial sequence ${\cal S}_1=S$ iteratively,
we obtain sequences $\{L, LS, LSL, LSLLS, \cdots \}$.
The $i$th sequence ${\cal S}_i$ is composed of $F_i$ letters,
where $F_i$ is the Fibonacci number.
We deal with the tightbinding model with the total number of sites $N=F_{i}$ and
the hopping integral is given as $v_n=v_L (v_S)$
if the $n$th letter of the Fibonacci sequence ${\cal S}_{i}$ is $L$ ($S$).

In the paper, we study the linear optical response for the tightbinding model on the Fibonacci chain.
The optical conductivity is given as $\sigma(\omega)=J(\omega)/{\cal E}(\omega)$,
where $J(\omega)$ and ${\cal E}(\omega)$ are
the Fourier components of the current $J(t)$ and electric field ${\cal E}(t)$. 
The current operator is split into two parts as
$\hat{J}(t) =\hat{j}_1+\hat{j}_2(t)$, with
\begin{align}
  \hat{j}_1=&-iq \sum_n v_n L_n (\hc_n^{\dag} \hc_{n+1} - \hc_{n+1}^{\dag} \hc_n),\\
\hat{j}_2(t)=&- q^2 \sum_n v_n L_n^2 (\hc_n^{\dag} \hc_{n+1} + \hc_{n+1}^{\dag} \hc_n) A(t).
\end{align}
By means of the Kubo formula, the optical conductivity is expressed as 
\begin{align}
\label{eq:sigma}
\sigma(\omega)
&=\frac{1}{N}\sum_{a,b} 
    \frac{f(E_b)-f(E_a)}{i(\omega+i\delta)} 
    \frac{| \bra{b}\hat{j} \ket{a}|^2}{\omega-(E_b-E_a)+i\delta} \nonumber\\
    &-\frac{1}{N}\sum_a
    \frac{q^2f(E_a)}{i(\omega+i\delta)}
    \bra{a}
    \sum_n v_n L_n^2
    (\hat{c}^{\dag}_n\hat{c}_{n+1}+\hat{c}^{\dag}_{n+1}\hat{c}_n)
    \ket{a},
\end{align}
where $\ket{a}$ is a single electron eigenstate of
the model with $A(t)=0$,
$E_a$ is the corresponding energy,
$f(x)$ is the Fermi distribution function, and $\delta$ is infinitesimal.
The first term represents the optical transition, and the other is 
the so-called Drude part.
For the sake of simplicity, we set $L_n=1$ in order to focus on the effects
of the Fibonacci structure in the hopping
integrals~\cite{PhysRevB.70.144207, PhysRevLett.60.1672}.

In the following, we mainly consider the system with $N=F_{19}=4181$
under the periodic boundary condition.
This system is large enough to take the quasiperiodic structure into account.
In fact, we have confirmed that
the obtained results are essentially the same as those with $N=17711$.
Here, setting the Fermi energy at $E=0$ and $v_L$ as the unit of energy,
we discuss how the quasiperiodic structure affects the optical linear response.

\section{RESULT}\label{sec:result}
We study the optical response in the tightbinding model with the Fibonacci structure.
First, we briefly discuss the one-particle states in the model
without the external electric field.
 \begin{figure}[htb]
  \centering
  \includegraphics[width=\linewidth]{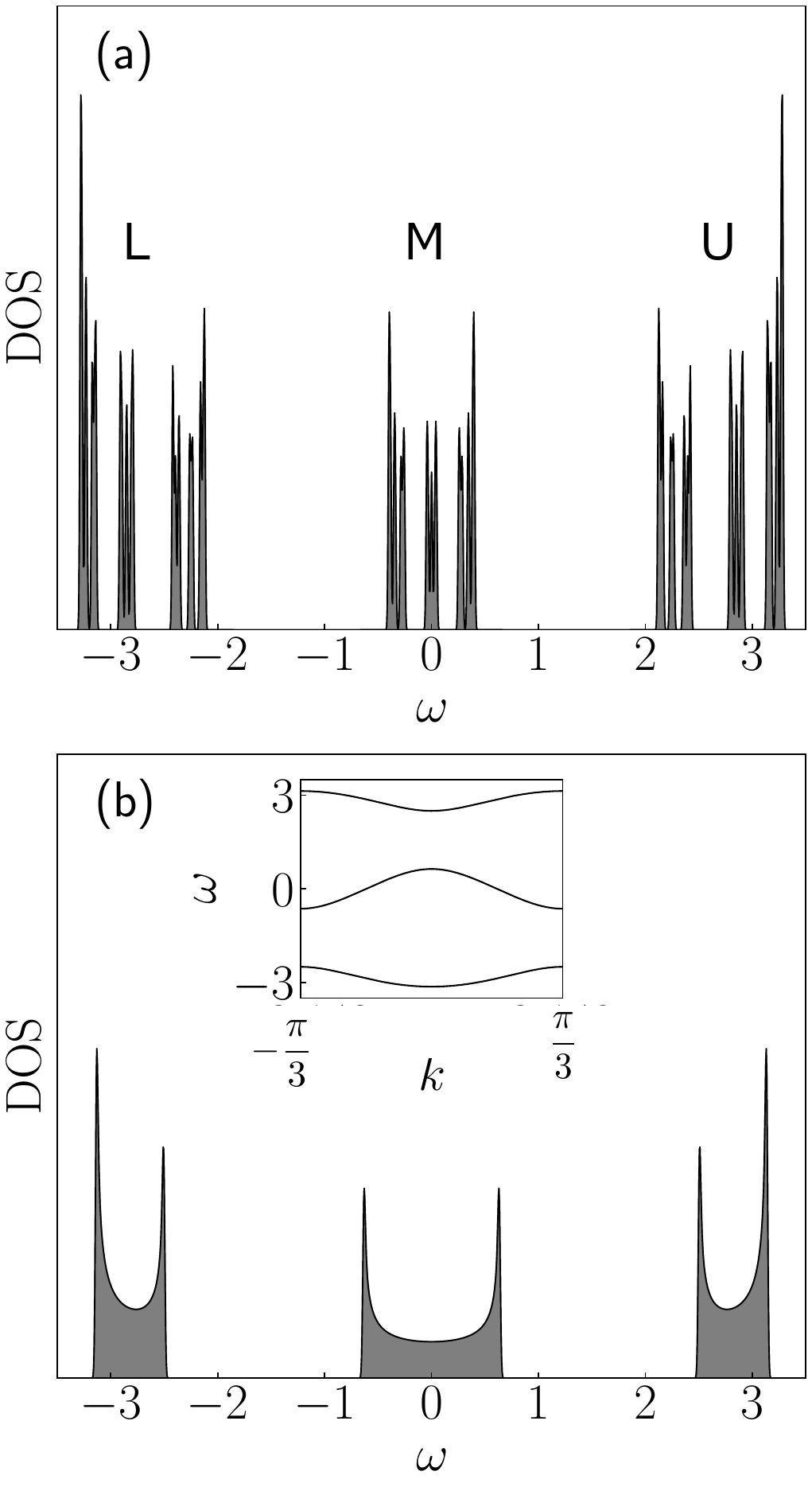}
  \caption{Density of states in the tightbinding model
    on the Fibonacci chain (a) and
    the approximant (b) with $v_S=2.5$ and $A(t)=0$.
    Inset in (b) shows the dispersion relation of
    the model for the approximant. The vertical axes are in an arbitrary unit.
  }
  \label{fig:fig1}
\end{figure}
Figure~\ref{fig:fig1}(a) shows the density of states (DOS)
for the tightbinding model on the Fibonacci chain with $v_S=2.5$.
Many delta-function peaks appear in the DOS,
which is known to be purely singular continuous
like the Cantor set ~\cite{KohmotoKadanoff, PhysRevB.34.2041,Kohmoto-critical1, Suto:1989uu}.
To plot a delta-function peak in DOS, we practically use the Gaussian with a small width in Figure~\ref{fig:fig1}. We also use the same strategy to plot delta-function peaks in the following. 
The energy levels are roughly classified
into lower (L), middle (M), and upper (U) bands [see Fig.~\ref{fig:fig1}].
When one focuses on the M band, there exist three smaller bands,
suggesting the nested structure in the DOS~\cite{doi:10.1143/JPSJ.55.3709}.
In fact, the DOS scaled by $R$ is in a good agreement with the original one,
where $R=E^U_{max}/E^M_{max}$ and $E^\alpha_{max}$ is the maximum energy
in the $\alpha(=U, M)$ band. 
It is also known that single-particle states of this model are critical
and the decay for each wave function is
slower than exponential one~\cite{Kohmoto-critical1, PhysRevB.40.7413, PhysRevB.34.563, KohmotoKadanoff, MaceJagannathan}.
In contrast to the Fibonacci case,
the smooth DOS should appear in the periodic system.
For comparison, we consider the tightbinding model on the approximant,
where the shorter Fibonacci sequence ${\cal S}_4(=LSL)$ is periodically arranged.
This is the minimal approximant with three bands in common with the Fibonacci chain,
as shown in Fig.~\ref{fig:fig1}(b).
Since each single-particle state in the approximant is characterized by the momentum,
its wave function is spatially extended,
in contrast to the Fibonacci case.
When one considers the approximant with the longer sequence,
the corresponding DOS approaches the Fibonacci one
and critical behavior should appear in the wave function.

\begin{figure}[htb]
  \centering
  \includegraphics[width=\linewidth]{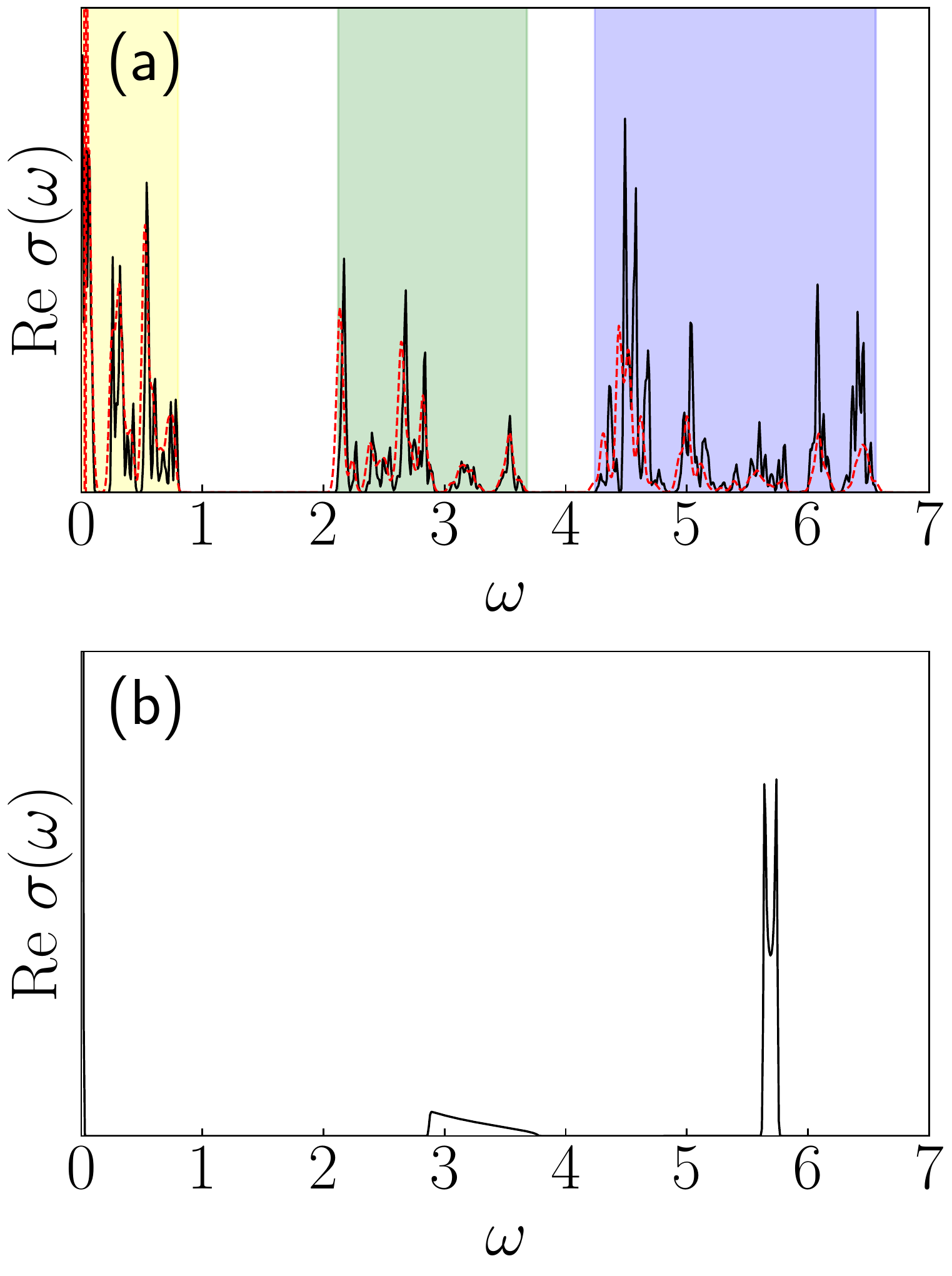}
  \caption{(Color online) Real part of the optical conductivity
  for the Fibonacci chain (a) and approximant (b) with $v_S=2.5$. The dashed line in (a) shows the optical conductivity scaled by $R$.
  The areas shaded in yellow, green, and blue represent different types of excitations, see the text for details. The vertical axes are in an arbitrary unit.
  }
  \label{fig2}
\end{figure}
Now we discuss the optical conductivity in these tightbinding models
with distinct behaviors in the one-particle states.
Figure~\ref{fig2} shows the real part of the optical conductivities
in the systems with $v_S=2.5$.
It is clarified that several peaks appear in the Fibonacci case, while
few structures appear in the approximant.
The peaks in the former can be explained by
the energy spectrum in the DOS,
where there exist L, M, and U bands [see Fig.~\ref{fig:fig1}(a)].  
When the half-filled model is considered,
the optical response is
categorized into three types of excitations;
those within the M band, between the M band and the other bands,
and between the L and U bands.
In fact, these excitations are clearly separated,
which are shown as the yellow, green, and blue areas in Fig.~\ref{fig2}(a).
The areas shaded in yellow, green, and blue represent the excitations within the M band, between the M band and the other bands, and  between the L band and the U band, respectively.
Furthermore, a large intensity in the optical conductivity appears,
which corresponds to the optical transition
between the energy levels with the large DOS.
By contrast, a rather simple structure appears
in the approximant with the similar DOS, as shown in Fig.~\ref{fig2}(b).
It is clarified that
no peak structure appears in the low frequency region $0<\omega<1$ and
in higher frequency region, peaks are not widely distributed, in contrast to the Fibonacci case.
Furthermore, the gap structure in the optical conductivity
is not directly related to the DOS.
Thus, the DOS is not enough to explain the optical response for the approximants.
This originates from the existence of the translational symmetry in the approximant,
which leads to the selection rule for the optical response.
Namely, the optical transition between distinct wave numbers is forbidden
[see Fig.~\ref{fig2}(b)].
Only the interband excitations contribute to the optical transitions with finite frequencies. 
Namely, the transition within the M band is absent, although there is a signal from the second term of eq. ~\eqref{eq:sigma} at $\omega=0$ , i.e. the Durde peak.

\begin{figure}[htb]
  \centering
  \includegraphics[width=\linewidth]{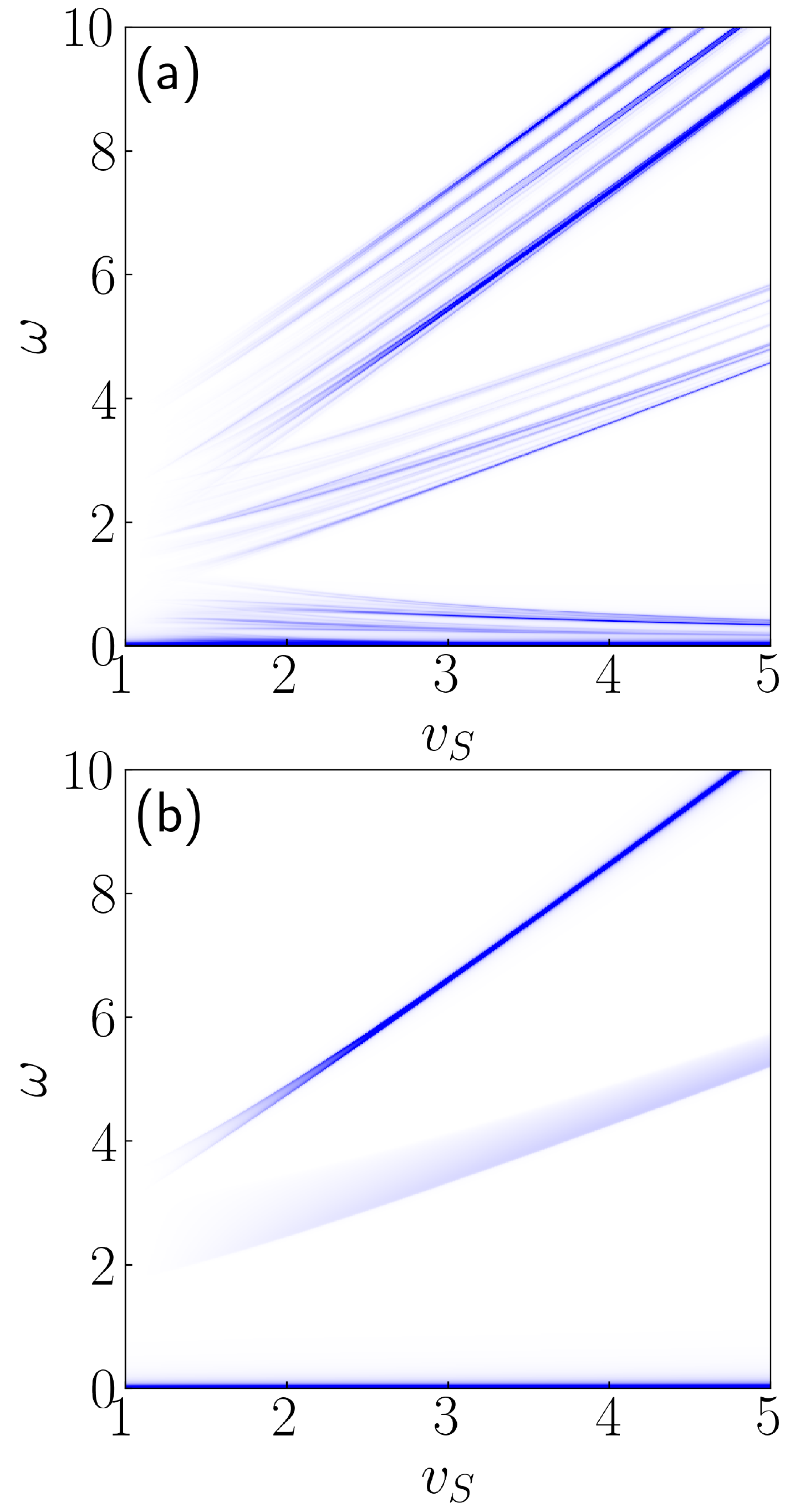}
  \caption{Real part of the optical conductivity as a function of $v_S$ for the Fibonacci chain (a) and
    the approximant (b). The color scale is in an arbitrary unit.}
  \label{fig3}
\end{figure}
The $v_S$ dependent conductivity is shown in Fig.~\ref{fig3}.
When $v_S\lesssim 2$, the energy ranges for intra- and interband excitations
overlap and their intensities are rather small.
Therefore, it is hard to see the characteristic feature
of the Fibonacci chain.
By contrast, the large $v_S$ leads to different behavior in the optical conductivity.
In the case, two interband excitations are almost
proportional to $2v_S$ and $v_S$,
which allows us to distinguish these excitations in the optical conductivity.
With the large $v_S$, the intensities of the optical conductivity are clear both in Fibonacci and approximant cases.
On the other hand, qualitatively distinct behavior
appears in each system:
in the Fibonacci case, several peak structures in the intra- and interband exciations
appear even when $v_S$ is large.

\begin{figure}[htb]
\includegraphics[width=\linewidth]{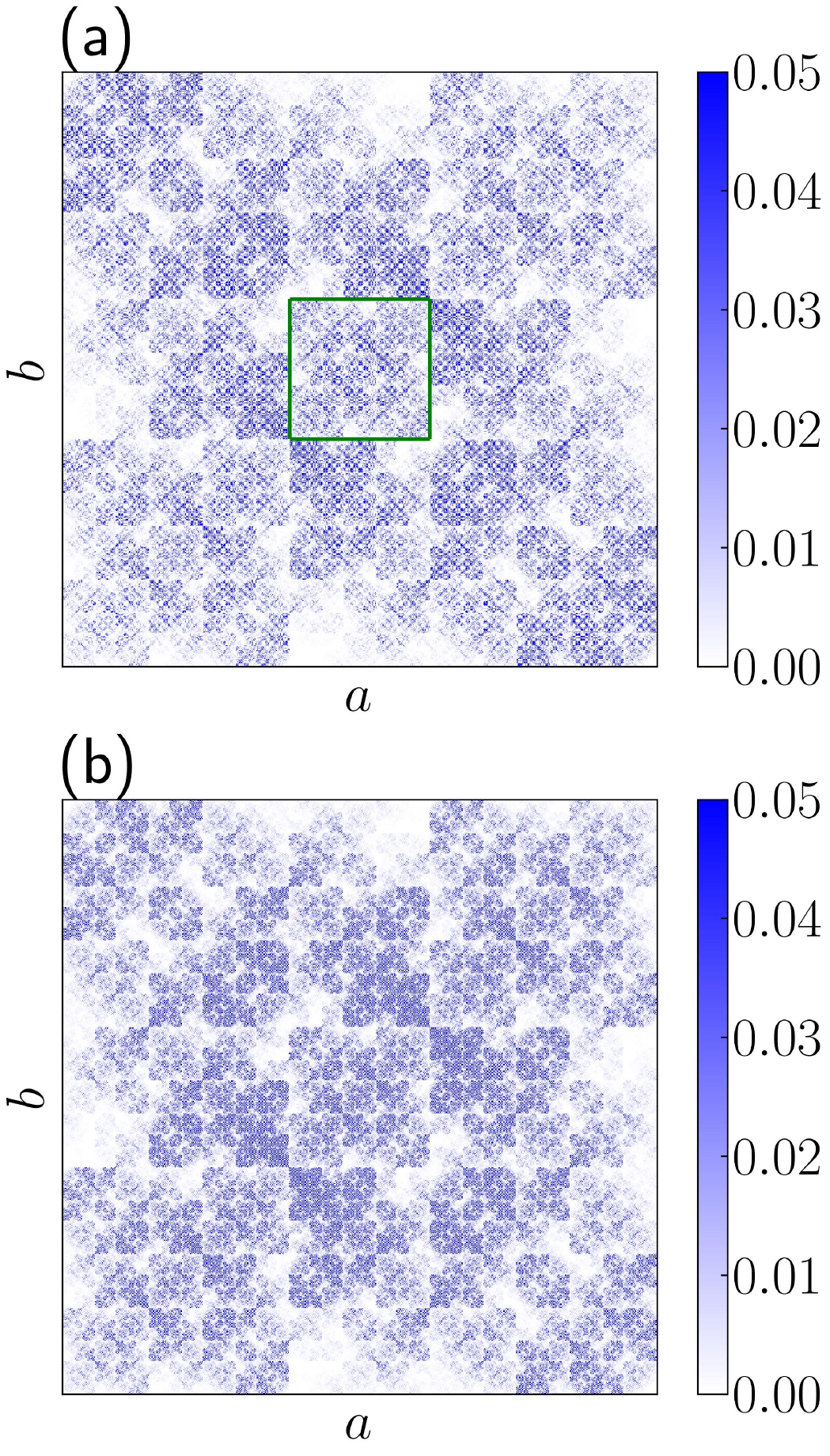}
\caption{(a) Matrix elements of the current operator $|\braket{b|\hat{j}|a}|$ for the Fibonacci chain.
(b) Magnified figures of the rectangular part in (a).
}
\label{fig4}
 \end{figure}

In order to obtain further insight into many spiky behavior
in the optical conductivity for the Fibonacci chain,
we focus on the matrix elements of the current operator $\bra{b}\hat{j}\ket{a}$
in eq. (\ref{eq:sigma}),
where the indices $a$ and $b$ are listed in ascending order by the eigenenergies.
We note that no degeneracy appears in the energy levels
and the matrix elements are uniquely determined.
Figure~\ref{fig4} shows the absolute value of the matrix elements of the current operator for the Fibonacci chain.
We find the dense structure in the matrix elements
although some of them is invisible in the figure.
This implies that
the optical transition between an arbitrary pair of states is allowed.
This should be consistent with the fact that
there exists no translational symmetry and
each eigenstate in the tightbinding model is critical.
By this reason, such optical properties realized in the quasiperiodic systems 
do not depend on the details of band structure. 
In the tightbinding model on the Fibonacci sequence ${\cal S}_n$,
the M band is composed of the $F_{n-3}$ energy levels,
and the others are composed of the $F_{n-2}$ levels~\cite{PhysRevA.35.1467}.
The matrix elements should be classified into nine groups,
and each has the dense structure.
One of the most interesting points is that self-similar behavior appears.
When one focuses on the matrix elements for the M band,
which is marked by the green square in Fig.~\ref{fig4}(a),
its pattern is similar to the original one.
This self-similar pattern enriches the peak structure in the optical conductivity.
Namely, some intraband excitations in the M band are related to the interband excitation
between L and U bands.
In the sense, the self-similar behavior in the matrix elements are reflected in the optical conductivity. 
In fact, the optical conductivity scaled by $R$
is in a good agreement with the original one,
as shown in Fig.~\ref{fig2}(a).
This is in contrast to the conventional periodic system, where
the optical transition is restricted between the states with the same momentum
(optical selection rule).

%
%

Up to now, we have discussed the optical responses
for the tightbinding models on the Fibonacci and its approximant,
and have clarified that the structure in the current operator leads to
spiky and self-similar behavior in the optical conductivity
for the Fibonacci system.
It is naively expected that disordered systems should have 
a dense structure in the matrix elements.
Therefore, it should be instructive to clarify the effect of the disorders for the optical response
in the Fibonacci system.
In this study, the bond disorder for the Fibonacci sequence is introduced.
A disorder is created by picking a random site $i$ and swapping its connecting hoppings $v_i$ and $v_{i-1}$;
$v_i \leftrightarrow v_{i-1}$.
We note that the tightbinding model with large disorders
is not reduced to the random-hopping model
since the numbers of $v_L$ and $v_S$ are conserved.
Here, we clarify how the self-similar structure in the optical response
is affected by the introduction of the disorders.
\begin{figure}[H]
\centering
\includegraphics[width=\linewidth]{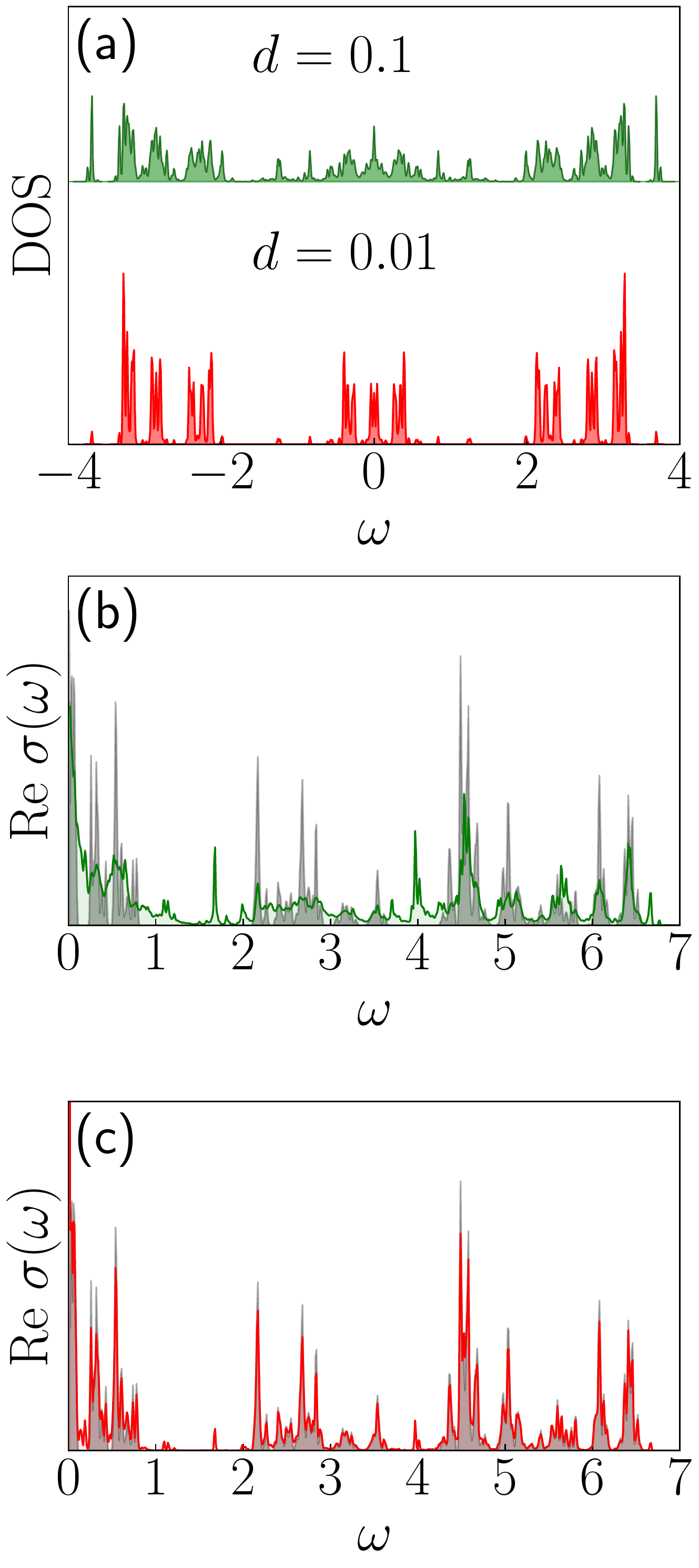}
\caption{(a) DOS of the disordered system with $v_S=2.5$
  when $d=0.01$ and $d=0.1$. The real part of the optical conductivity for $v_S=2.5$ when (b)$d=0.1$, (c)$d=0.01$. 
  The areas shaded in grey depict the optical conductivity of the Fibonacci chain as shown in Fig. ~\ref{fig2}(a).
   The vertical axes are in an arbitrary unit.
  }
\label{fig5}
\end{figure}
Figure ~\ref{fig5}(a) shows the DOS for the Fibonacci chains with $d=0.01$, and $0.1$,
where $d(=m/N)$ is the density of the disorders and $m$ is the number of swap operations.
The results are evaluated by means of, at least, a thousand independent random samples
and standard deviations are invisible in the figure.
When the disorders are introduced in the one-dimensional chain,
it may be regarded as the system composed of
Fibonacci chains with finite length.
In the case, the average of the length is roughly given by $1/d$ and
thereby the characteristic feature of the Fibonacci chain still remains when $d<0.01$.
Increasing the disorders, the singular continuous spectrum
inherent in the Fibonacci chain smears
and the spectrum have the intensity in the broader energy range,
as shown in Fig.~\ref{fig5}(a).
In Fig.~\ref{fig5}(b)(c), we show the optical conductivity in the system
with $d=0.01$, and $0.1$. 
We find that, increasing $d$, the peak structures which is clear
in the Fibonacci case smear {\it eg.} at $\omega\approx 0.6, 2.2, 2.7, 4.5, 6.0$.
Some peaks instead emerge with increasing $d$,
{\it eg.} at $\omega \approx 1.7, 4.0$.
  \begin{figure}[htb]
\centering
   \includegraphics[width=\linewidth]{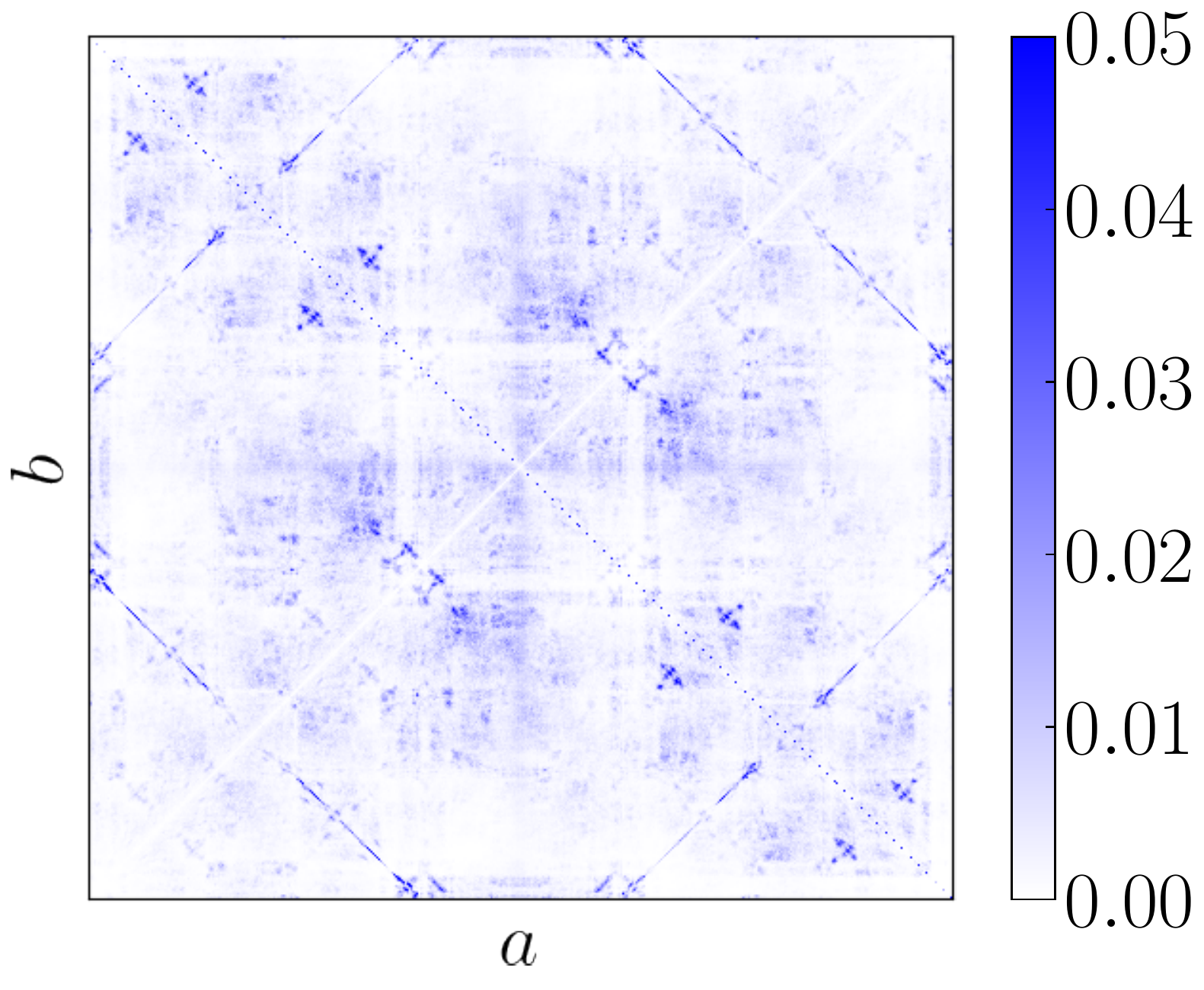}
   \caption{The matrix elements of the current operator $|\braket{b|\hat{j}|a}|$
      for the disordered Fibonacci chain with $v_S=2.5$ and $d = 0.1$. }
 \label{fig6}
  \end{figure}
This makes the self-similar structure in the conductivity, which is clarified by rescaling the frequency, less likely  with increasing $d$.
This is consistent with the fact that self-similar pattern
smears in the matrix elements of the current operator,
as shown in Fig.~\ref{fig6}.

\section{CONCLUSION}
We have investigated the optical response for the one-dimensional tightbinding model on the Fibonacci chain. 
We have calculated the optical conductivity in terms of the Kubo formula
and have clarified that it reflects the singular continuous DOS inherent in the Fibonacci chain.
This is contrast to those for the approximants,
where the optical transition is restricted 
due to the existence of the translational symmetry.
By examining the matrix elements of the current operator carefully,
we have found the self-similar structure in the optical conductivity,
where it is well scaled by the value $R$.
We have also discussed the effects of disorders in the Fibonacci chain and clarified that
the self-similar structures in optical conductivity smear.
From these analyses, we conclude that the self-similar feature observed in the optical conductivity
as well as the DOS originates from
the quasiperiodic structure inherent in the Fibonacci sequence.
Since the Fibonacci lattice can be realized for cold atoms on the optical lattice~\cite{PhysRevA.92.063426} and for photons in a photonic waveguide array~\cite{PhysRevB.91.064201}, these systems are potential playgrounds for our theoretical prediction.  In particular, for the cold atom systems, one can effectively realize the electric field ~\cite{Jotzu2014}. 
Our results suggest that physical observables such as optical conductivity can show peculiar structures originating from quasiperiodic lattice structures.
This result shall stimulate further experimental and theoretical studies to explore characteristic self-similar structures in optical conductivity in various quasiperiodic systems.
To undertand the origin and the generality of such structures,  detailed analyses of wave functions and/or various types of quasi-periodic systems are required. 
Furthermore, nonlinear optical responses have been actively studied in the context of condensed matter physics.
In future, it is also important to clarify how quasiperiodic structures affect nonlinear optical properties.

\acknowledgements
Parts of the numerical calculations are performed
in the supercomputing systems in ISSP, the University of Tokyo.
This work is supported by Grant-in-Aid for Scientific Research from JSPS,
KAKENHI Grant Nos. JP20K14412, JP20H05265, JP21H05017 (Y.M.) and
JP19H05821, JP18K04678, JP17K05536 (A.K.),
and JST CREST Grant No. JPMJCR1901 (Y.M.). \\

\bibliographystyle{apsrev4-1}
\bibliography{Master.bib}

\end{document}